\documentclass[10pt,conference]{IEEEtran}

\usepackage{amsmath,amssymb,amsfonts} 
\usepackage{graphicx} 
\usepackage{cite} 
\usepackage{hyperref} 
\usepackage{algorithmic} 
\usepackage{textcomp} 
\usepackage{xcolor} 
\usepackage{todonotes}
\usepackage{balance}
\usepackage{dsfont}
\usepackage{xspace}

\usepackage[normalem]{ulem} 

\usepackage{ifthen}
\usepackage{amssymb}
\usepackage{amsfonts}
\newboolean{showcomments}
\setboolean{showcomments}{true} 
\ifthenelse{\boolean{showcomments}}
  {\newcommand{\nb}[2]{
    \fcolorbox{gray}{yellow}{\bfseries\sffamily\scriptsize#1}
    {\sf\small$\blacktriangleright$\textit{#2}$\blacktriangleleft$}
   }
   
  }
  {\newcommand{\nb}[2]{}
   
  }

\usepackage{orcidlink}

\title{Measuring and Analyzing Resilience Metrics in Self-Adaptive Systems}
\title{Calculating and Analyzing Resilience Metrics in Self-Adaptive Systems}

\title{Analyzing Resilience to Understand Antifragility}
\title{\textsc{ResMetric}: Analyzing Resilience to Enable Research on Antifragility}

\author{
    \IEEEauthorblockN{Ferdinand Koenig\IEEEauthorrefmark{1}\orcidlink{0000-0003-3311-7355}, Marc Carwehl\IEEEauthorrefmark{1}\orcidlink{0000-0003-0631-365X}, Calum Imrie\IEEEauthorrefmark{2}\orcidlink{0009-0004-3198-9226}}
    \IEEEauthorblockA{\IEEEauthorrefmark{1}Institut für Informatik, Humboldt Universität zu Berlin
    , Germany\\
    Email: \{ferdinand.koenig, m.carwehl\}@hu-berlin.de}
    \IEEEauthorblockA{\IEEEauthorrefmark{2}Department of Computer Science, University of York, United Kingdom\\
    }
}

\newcommand{\proposal}{\textsc{ResMetric}\xspace}

\begin{document}
\maketitle

\begin{abstract}
A key feature in self-adaptive systems is resilience, which is an ongoing research topic. 
Recently, the community started to explore antifragility, which describes the improvement of resilience over time.
While there are model-agnostic resilience metrics, there is currently no out-of-the-box tool for researchers and practitioners to determine to which degree their system is resilient. 
To facilitate research on antifragility, we present \textsc{ResMetric}, a model-agnostic tool that calculates and visualizes various resilience metrics based on the quality of service over time. With \textsc{ResMetric}, researchers can evaluate their definition of resilience and antifragility. 
This paper highlights how \textsc{ResMetric} can be employed by demonstrating its use in a case study on gas detection. 

\end{abstract}


\IEEEpeerreviewmaketitle

\section{Introduction}
Research on self-adaptive systems (SASs) has largely focused on improving systems' resilience towards uncertainty in the past~\cite{almeida2011benchmarking, burton_et_al:DagRep.14.4.142}. To this end, many techniques exist that use, e.g., formal verification or automated reasoning to adapt a system's architecture~\cite{mahdavi2017systematic} or behavior~\cite{BSN}. Multiple quality metrics have been used to evaluate the effectiveness of such adaptions across a wide range of systems from many domains~\cite{wong2022self}. For instance, for a server system like ZNN~\cite{cheng2009evaluating} the aim might be to improve response time, whereas in a robotic system~\cite{carwehl2024formal} the stakeholders might value the probability of a successful mission. Since different quality metrics need to be used for different systems, there is also no consensus on how to best measure resilience for any system. In contrast, an adequate resilience metric needs to be chosen w.r.t. the suitable quality metric, the system under consideration, and the properties that stakeholders want to focus on. Lately, research has also investigated how we cannot only achieve resilience in SASs, but how we can take the next step towards \emph{antifragility}~\cite{grassiConceptualArchitecturalCharacterization2024}. Discussions, e.g., in a recent Dagstuhl seminar~\cite{burton_et_al:DagRep.14.4.142}, have centered around antifragility as the gain in resilience over time, i.e., a system that improves its resilience over time and across shocks, may be considered antifragile.

Due to the trend towards antifragility, choosing the ``correct'', that is, most fitting, resilience metric for a system becomes ever more important. To this end, we present \textsc{ResMetric}, a tool that computes a vast range of resilience metrics
, visualizes the resilience over time, and, therefore, enables meaningful discussions about antifragility. Additionally, \proposal implements the change in resilience, which has been discussed as a naive notion of antifragility. As a python package, \textsc{ResMetric} can be used straight out-of-the-box with many existing resilience metrics, such as area under the curve (AUC), robustness, or recovery rate. Finally, \textsc{ResMetric} is model-agnostic and requires as input only the system's performance over time as a \emph{JSON} file or a Plotly scatter plot. \proposal aims to assist researchers to:
\begin{itemize}
    \item evaluate how different systems suffer from uncertainty, 
    \item evaluate how different adaptation strategies help a system to react to uncertainty, 
    \item classify different disruptions by analyzing their effect on the system, and 
    \item determine how the system deals with repeated disruptions of the same kind. 
\end{itemize}

The remainder of this paper is structured as follows: Section~\ref{sec:background} provides background on resilience. 
Afterwards, we present the metrics implemented in \textsc{ResMetric} in Section~\ref{sec:approach} before demonstrating calculating and choosing resilience metrics for a gas detection system from the literature in Section~\ref{sec:demonstration}. We conclude the paper in Section~\ref{sec:conclusion} after reviewing related work in Section~\ref{sec:related}.

\begin{figure}[tb]
    \centering
    \includegraphics[width=0.9\linewidth]{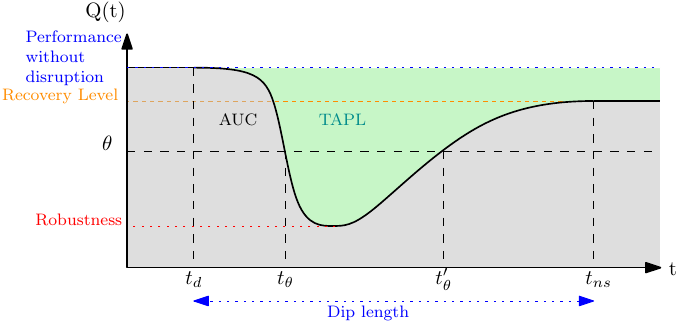}
    \caption{Model of a system's performance during and after a disruption.}
    \label{fig:dip}
    \vspace{-1em}
\end{figure}

\section{Background}\label{sec:background}
We consider that the system suffers from uncertainty which leads to disruptions. These disruptions can manifest in the system's performance. 
Fig.~\ref{fig:dip} visualizes one such \emph{dip} in the performance caused by a disruption and shows how the system recovers from it, e.g., through self-adaptation~\cite{grassiConceptualArchitecturalCharacterization2024}. For analysis, we consider time windows from $t_0$ to $t_1$, in which one dip occurs. The dip starts at $t_d$ and the system performance drops below a threshold $\theta$ at $t_\theta$. $t'_\theta$ represents the point in time at which the performance exceeds the threshold again before it settles at $t_{ns}$.

\section{\proposal}\label{sec:approach}

\begin{figure*}
    \centering
    \includegraphics[width=0.95\linewidth]{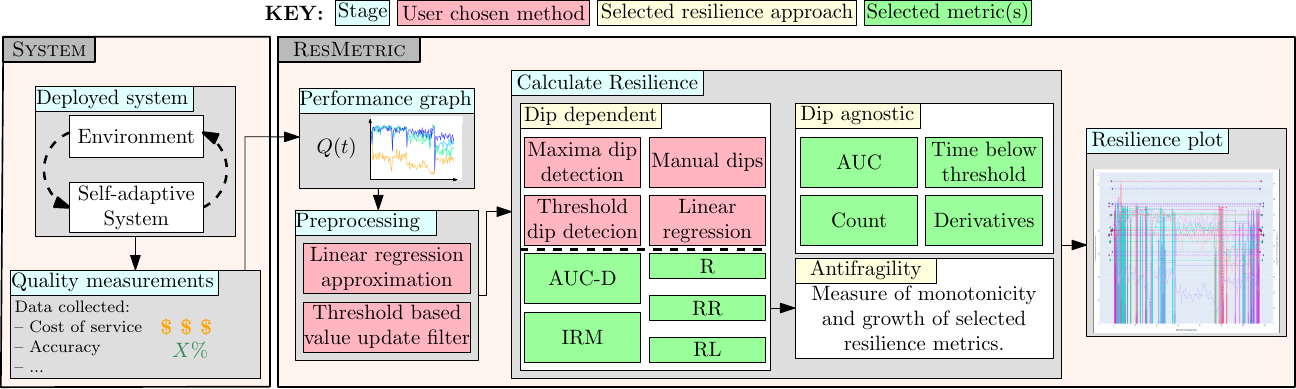}
    \caption{\textsc{ResMetric} tool architecture.}
    \label{fig:resmetric_diagram}
    \vspace{-1em}
\end{figure*}

This paper introduces \proposal\footnote{\url{github.com/ferdinand-koenig/resmetric}}\cite{koenig_2025_14726290}, a platform-independent Python package, which provides a command-line interface (CLI) for calculating and visualizing resilience-related metrics of systems. Fig.~\ref{fig:resmetric_diagram} visualizes the \textsc{ResMetric} process. The input is the system's performance over time, represented as a Plotly scatter plot or as JSON-serialized files (cf. Performance graph in Fig,~\ref{fig:resmetric_diagram}). The output is a Plotly graph visualizing the input along with resilience metrics (cf. Resilience plot in Fig.~\ref{fig:resmetric_diagram}). It can be saved as an HTML or JSON file or displayed as an interactive plot in the browser. \proposal assumes that the input data is normalized time series data in the interval $[0, 1]$. For a server-system~\cite{cheng2009evaluating}, this input data could be the ratio of handled requests over received requests. 

In general, such data can be noisy. Therefore, \proposal implements a threshold-based value update filter and linear regression approximation to pre-process the input data to remove noise. 

In the following, we present and discuss the different metrics implemented in \proposal that can be selected by the user. With \proposal, users can evaluate how resilient their system is and compare different adaptation strategies.

\subsection{Dip-Agnostic Resilience Metrics}
Some resilience metrics are agnostic to dips, i.e., we can calculate such metrics without explicit labels for dips (such as $t_d$ and $t_{ns}$). These metrics provide the advantage that labels do not have to be supplied as input or derived from the input data before analysis. 

\paragraph{Area under the curve} \emph{AUC} is one of the most popular of the dip-agnostic metrics~\cite{cimellaro2010}. It is determined by calculating the area under the performance graph, which will be the largest if the system performance is maximal. Fig.~\ref{fig:dip} visualizes AUC as a gray area. Since \proposal relies on discrete data, the trapezoidal rule is implemented to calculate AUC. To ensure comparability of AUC across differently sized time windows, we normalize AUC by the length of the time window. 

To provide a weighted moving average of AUC, \proposal additionally implements kernel functions that assign greater weight to more recent data points. To this end, the user can select between two kernels, one that implements exponential weight decay (where the half-time is user-set), and another one with inverse weighting, to specify how much the weight should be shifted. Using a uniform kernel, the traditional AUC is computed. Additionally, the user can also provide their own kernel function. 

AUC offers the advantage that no additional parameters have to be set and no further assumptions have to be made about the input data. 
However, being such a simple metric, AUC might be unable to focus on interesting aspects of the data. For instance, consider the following two dips: During the first dip, the performance drops very low but the system is able to restore performance very quickly; for the other dip, the performance is reduced only a bit, but the system takes tremendously longer to restore performance. The AUC for both dips might be the same. 

\paragraph{Thresholds} To address some of the disadvantages of AUC, threshold-based metrics have been discussed~\cite{burton_et_al:DagRep.14.4.142}. Similar to AUC, no labels for dips are required here, only a threshold $\theta$ that describes a desired minimum system performance. Using this threshold, we can calculate the amount of time that the system performance was below the threshold. This metric can be helpful when thresholds have been set by the stakeholders. \proposal also reports a count of the instances when the system is below the threshold.

\paragraph{Derivatives} Understanding how the quality performance changes over time is paramount for analyzing systems. Within the context of evaluating resilience derivatives can be utilized for sensitivity analysis for example. \proposal can provide the first and second derivative for each $Q(t)$ provided. 

\subsection{Dip-Dependent Metrics}
More sophisticated metrics require knowledge about dips. To this end, \proposal can automatically detect dips, using one of three available algorithms, or have this information provided by the user as input:

\noindent \textbf{Manual Dips: } If the user has already identified dips in the performance graph, both can be input to \proposal.

\noindent \textbf{Max Dips: } 
This algorithm finds the local maxima of the system performance and identifies samples between maxima as dips. This is the fastest detection algorithm available and does not require any additional inputs.

\noindent \textbf{Threshold Dips: }
In contrast, this algorithm relies on a threshold $\theta$ and identifies dips as samples that are below $\theta$. 

\noindent \textbf{Linear Regression Dips: }
While the previous two algorithms may have problems identifying steady states, this algorithm relies on linear regression to identify segments of steady states. To identify how many segments the series should be divided into, \proposal relies on Bayesian optimization. We recommend using this dip detection as the Bayesian optimization appears to provide the best results. However, it takes tremendously longer to compute than the previous two dip detection algorithms.

For 
our demonstration, this function used roughly 5 minutes per quality signal (136 data points each), compared to less than 1s for the other methods. 

\paragraph{AUC of dips}
With dips, \proposal can also calculate the AUC of the dip (\emph{AUC-D}), without the area during steady states, i.e., before and after the dip. AUC-D is the area divided by the time frame for the associated dip.

\paragraph{Robustness}
Robustness (\emph{R}) is used to refer to the extent to which a system maintains its functionality during and after a disruption~\cite{sansaviniEngineeringResilienceCritical2017}. Other terms have emerged, such as maximum impact level, drawdown, or maximum performance degradation. Essentially, they all describe the minimum level of performance that the system is able to provide. As such, \proposal calculates the minimum performance for a given time window. It is visualized in red in Fig.~\ref{fig:dip}. This metric can be used when the overall performance is impacted heavily by the minimum performance, such as in an autonomous vehicle where a sudden drop in sensor accuracy during critical manoeuvres, like obstacle avoidance, could compromise safety.

\paragraph{Recovery Rate}
Another dimension of a dip is its length~\cite{songSystemResilienceDistribution2022}.  
Fig.~\ref{fig:dip} visualizes dip length as a blue arrow along the time axis. \proposal calculates the recovery rate (\emph{RR}) as the reciprocal of dip length, i.e., a higher recovery rate indicates a faster performance recovery. This metric is particularly useful in scenarios where the time taken to recover performance is critical, such as in robotic service assistants resuming normal operations quickly after encountering a dip in performance.

\paragraph{Recovery Level}
However, not just the RR might be of interest, but also the performance that the system provides after the dip, i.e., in the next steady state. This is visualized in orange in Fig.~\ref{fig:dip}, and is known as the recovery level (\emph{RL}).
Francis and Bekera~\cite{francisMetricFrameworksResilience2014} defined \textit{adaptive capacity}, the fraction of performance in the subsequent steady state over the performance before the dip.
Sansavini~\cite{sansaviniEngineeringResilienceCritical2017} proposed \textit{recovery ability} as the ratio of how much the performance recovered over how much the performance deteriorated during the dip, extending adaptive capacity to include information about robustness. 
\proposal calculates both the adaptive capacity as well as recovery ability. These metrics are particularly relevant in systems where maintaining or improving performance after a dip is crucial. For instance, in cloud-based service systems, adaptive capacity can be used to evaluate how well the system restores service quality after a server outage.

\paragraph{Integrated Resilience Metric}
Multiple proposals have been made to establish a rich metric that includes various aspects of system quality~\cite{francisMetricFrameworksResilience2014, mcdanielsFosteringResilienceExtreme2008, sansaviniEngineeringResilienceCritical2017}. Among them GR~\cite{sansaviniEngineeringResilienceCritical2017}, which relies on rapidity (RAPI) and time-averaged performance loss (TAPL). 
RAPI, on the one hand, resembles the fraction of the average slope of the recovery phase over the disruption phases, i.e., it specifies if the recovery phase was steeper than the disruption phase. 
TAPL, on the other hand, resembles the performance loss caused by the disruption. Assuming that without the disruption the performance would be constant, TAPL calculates the area between this assumed constant performance and the actual performance curve (green area in Fig.~\ref{fig:dip}), normalized by the duration of the dip.
However, we found a limitation in the definition of GR in~\cite{sansaviniEngineeringResilienceCritical2017}. If we assume that the system can have a better performance after the dip (i.e., after an adaptation) than before, then TAPL will be negative because it represents a directed area.

\proposal implements an integrated resilience metric (\emph{IRM}) that is similar to GR, but acknowledges that the system might improve its performance over time. To this end, we increment TAPL before using it in IRM. This ensures that TAPL is always positive. Similar to the literature, IRM calculates the product of the following metrics: Robustness, Rapidity, (TAPL+1)$^{-1}$, and Recovery.

However, we want to point out that IRM implicitly assumes a weighting from the metrics it relies on. This might be a problem since the metrics are defined over different domains. Robustness, Recovery, and TAPL, on the one hand, can never exceed 1. Rapidity, on the other hand, can be any real value. This implicit weighting is already present in GR.

\subsection{Towards Antifragility}
Recent discussions in the community proposed further research on antifragility~\cite{burton_et_al:DagRep.14.4.142}. 
Antifragility is a system's ability to improve its response to disruptions, i.e., to improve its resilience over time~\cite{grassiConceptualArchitecturalCharacterization2024}.
According to Grassi et al.~\cite{grassiConceptualArchitecturalCharacterization2024}, antifragility is achieved if resilience monotonically increases, but discussions in a recent Dagstuhl seminar~\cite{burton_et_al:DagRep.14.4.142} highlighted that we might want to compare systems regarding their antifragility, i.e., to determine which system is \emph{more antifragile}.  

\proposal computes the degree of antifragility over a resilience metric $u$ as $\alpha_u$ in response to disruptions using the average change (increase or decrease) in resilience. 
To this end, \proposal defines antifragility such that: 
When $\alpha_u = 0$, the resilience metric is strictly monotonically decreasing, indicating fragility. If $0 < \alpha_u < 1$, $\alpha_u$ serves as a membership function for monotonicity, reflecting the ratio of upward to downward changes in resilience. When $\alpha_u \geq 1$, it signifies a monotonically increasing resilience metric $u$, with values calculated as the average rate of improvement, incremented by one. A higher $\alpha_u$ indicates a faster rate of improvement in the resilience of the system. \proposal also provides the mean antifragilty, \(\alpha_{\bar{u}}\).

\subsection{Extendability}
The functions implemented for dip detections, resilience metrics, and antifragility are designed to be modular. This allows users to experiment with specific aspects and expand the current repertoire. Examples include implementing additional resilience metrics and examining approaches for antifragility. 

We recommend as guidance for users wanting to extend \textsc{ResMetric} to first look at the \texttt{metrics.py} file to give clarity on function implementation. Specifically to better understand how to incorporate the existing methods for their extensions (e.g. dip detection). We then suggest the user to examine \texttt{plot.py} for a particular setup (e.g. dip-agnostic selecting AUC as the only resilience metric) for clarity and inspiration for visualization. 

\section{Verifying gas delivery}
\label{sec:demonstration}
To demonstrate how to use \textsc{ResMetric} the artifact contains an example case study. Specifically, the self-adaptive system for lifelong machine learning utilised for classifying gasses in the pipes as part of a gas delivery managing system. We use existing models from the literature and introduce a new model to show how they can be compared using \textsc{ResMetric}.

\subsection{Self-adaptive system for lifelong machine learning}
A gas delivery manager is concerned with ensuring that users are provided with the appropriate gas for their needs. This involves combining various chemicals to create the gas which is then routed to the user. However, there is uncertainty as to whether the intended gas was properly produced and therefore requires verification before being delivered to the user. Machine learning can be employed to classify the gasses, though data shifts can occur due to sensor degradation~\cite{vergara2012chemGasSensorDrift, Vergara2013gasSensorArrayData}. 

Gheibi and Weyns devised a self-adaptive system for a lifelong machine learning model, demonstrating its effectiveness in classifying gas in a pipeline~\cite{Gheibi2022-LifelongSelfAdapt}. The artifact contains the classification accuracy graph from this work as the performance quality input for \textsc{ResMetric}. While this graph is not included in their paper, it reflects the same reported results and can be accessed through their replication package\footnote{\url{https://people.cs.kuleuven.be/~danny.weyns/software/LLSAS/}}. Gheibi and Weyns used three support vector classifiers (SVC) for their investigation; an offline SVC to act as a baseline, an SVC which used the previous batch for online learning, and SVCs in the proposed lifelong learning loop (LLL). 

\begin{figure}
    \centering
    \includegraphics[width=0.85\linewidth]{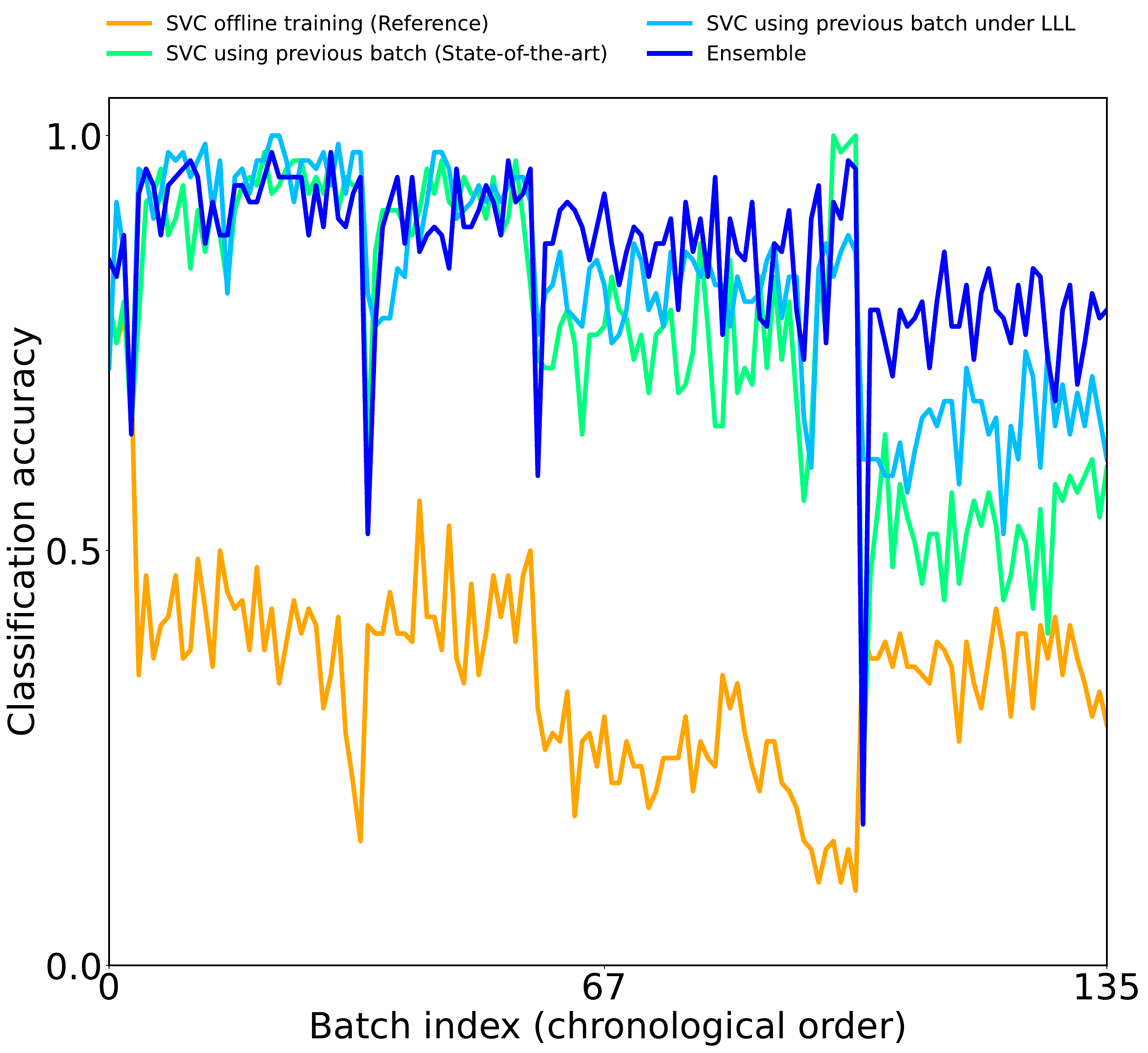}
    \caption{The performance quality of the four models investigated in the gas delivery case study.}
    \label{fig:performance_quality}
\end{figure}

We added a fourth curve by incorporating the classification performance of an ensemble learner, complementing the three existing performance curves, see Fig.~\ref{fig:performance_quality}. The ensemble learner is constructed via Gradient Boosting and is integrated within the MAPE-K loop of the lifelong self-adaptive system. We will now use \textsc{ResMetric} to compare the new model against the others mentioned with respect to resilience. 

\begin{figure}
    \centering
    \includegraphics[width=0.85\linewidth]{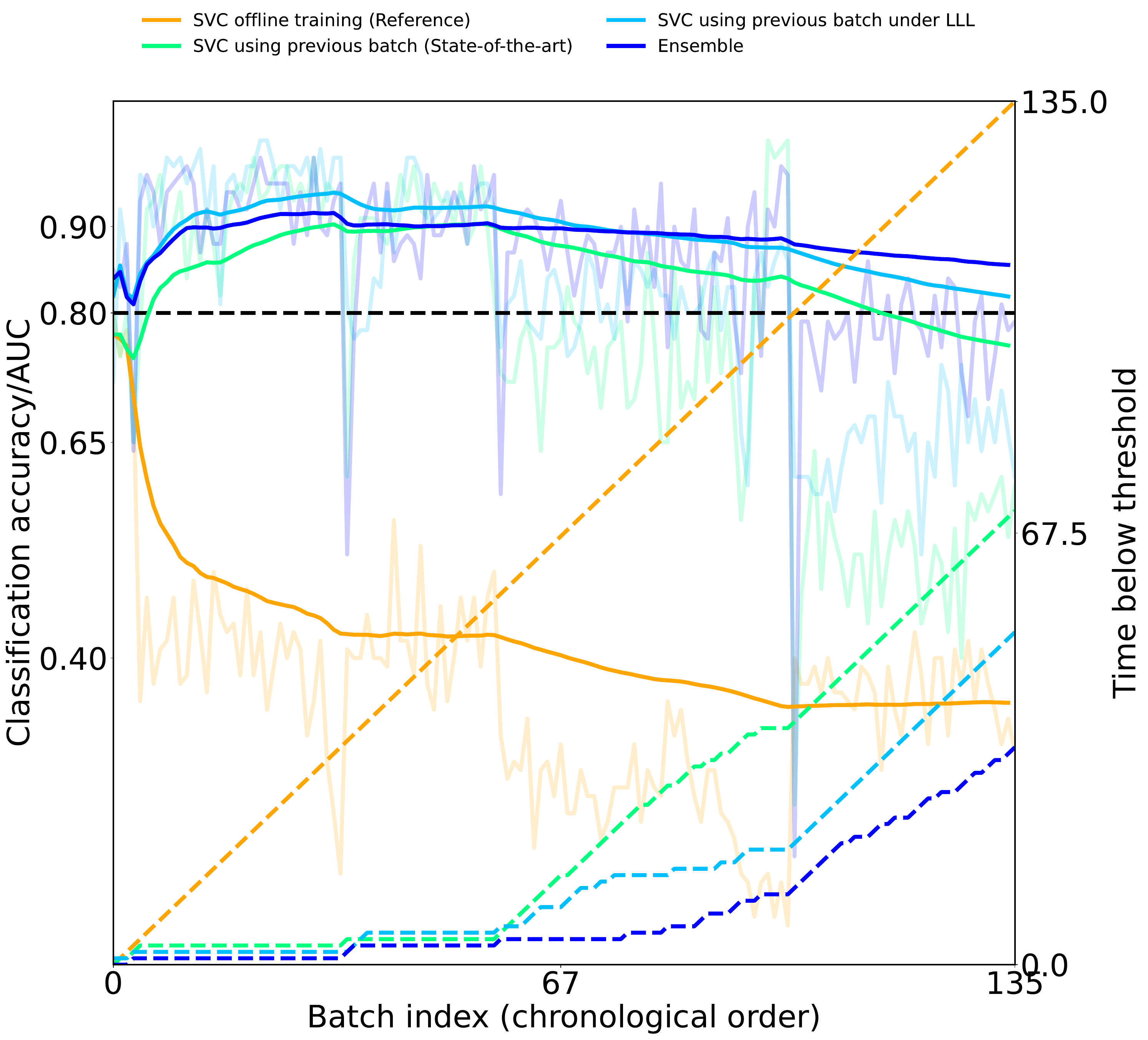}
    \caption{Resilience Metrics: Area Under Curve (AUC) (solid non-faded lines), and Time spent below threshold (dashed lines). The threshold is set at 0.8 (dashed black line). }
    \label{fig:example1}
\end{figure}

\begin{figure*}[tbh]
    \centering
    \includegraphics[width=0.85\linewidth]{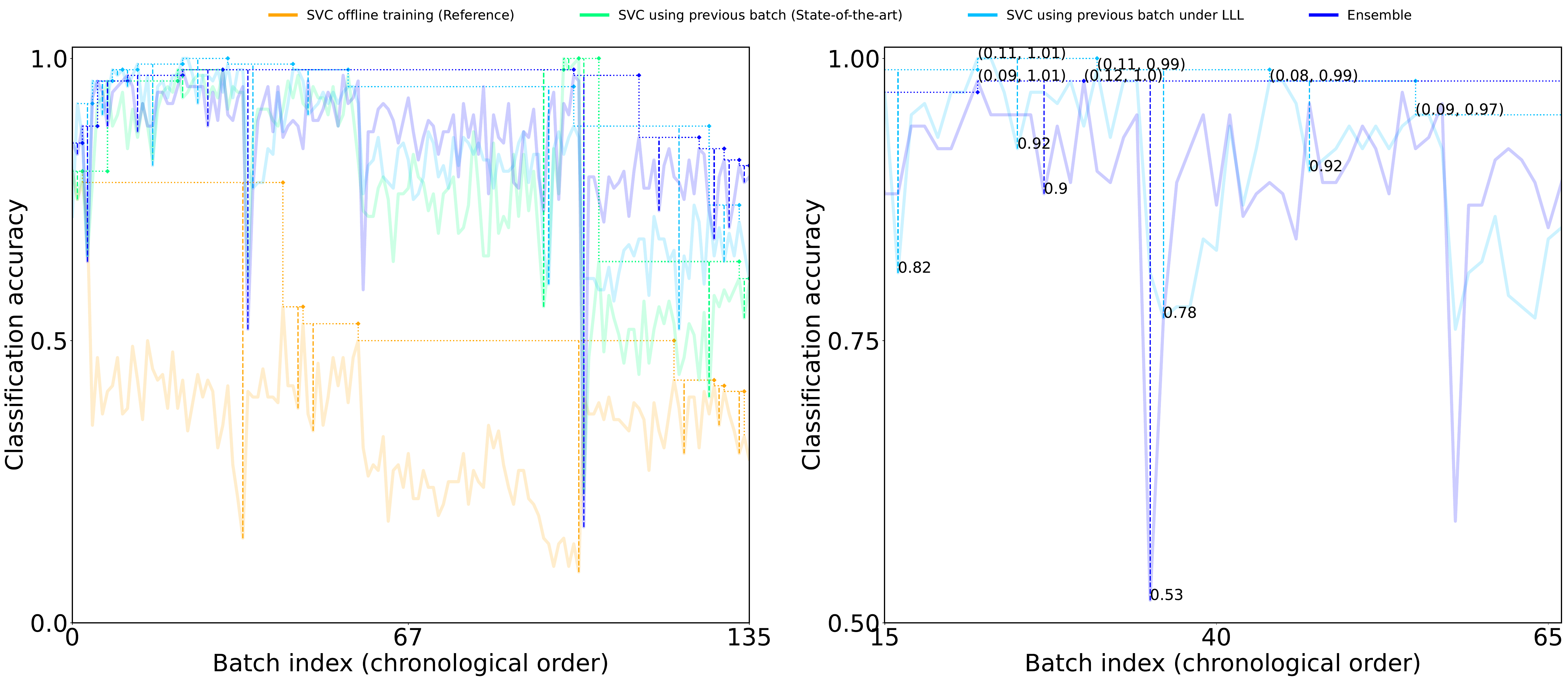}
    \caption{R, RR, and RL based on detected dips. Vertical dashed lines represent the maximum drop used for calculating, R, while the dotted lines and markers are for RR and RL. The plot on the right is a subselection of batches and methods to demonstrate the labeling. Robustness is at the maximum drops of dips, and the recovery rate/ability is labeled as a tuple.}
    \label{fig:example2}
\end{figure*}

\textsc{ResMetric} generates interactive plots, and when hovering over the traces in the plot, additional details such as exact values and metric names are displayed. This provides an exploration of the resilience metrics, and the interactive plots for this case study are available in the artifact. Further details on how to navigate the interactive plots can be found in the README in the artifact's repo.

\subsection{Dip-Agnostic Resilience Metrics}
We calculate dip-agnostic resilience metrics, which measure resilience without knowing the exact start and end points of disruptions. We look at two metrics: AUC, and the cumulative time the system performed below a given threshold. A uniform kernel was used for the AUC and the threshold was set to 80\% with respect to the classification accuracy. 

Fig.~\ref{fig:example1} contains the results. The SVC offline training performs worse over time and was always below the threshold. The other models consistently remained above 80\% classification accuracy until near the end of the batches.

\subsection{Robustness, Recovery Rate, and Recovery Level}
Robustness, recovery rate, and recovery level are three dip-dependent metrics. Here we use adaptive capacity for evaluating recovery level. The maxima-based dip-detection algorithm was chosen. The analysis, see Fig.~\ref{fig:example2}, includes plots highlighting the three metrics. The interactive plots also provide quantitative annotations which is demonstrated in Fig.~\ref{fig:example2} as well. We can observe, for example, that the ensemble and LLL systems have fairly similar scores, while the offline model has typically lower recovery rates.

\subsection{Linear Regression with Auto Segmentation}
\begin{figure}
    \centering
    \includegraphics[width=0.85\linewidth]{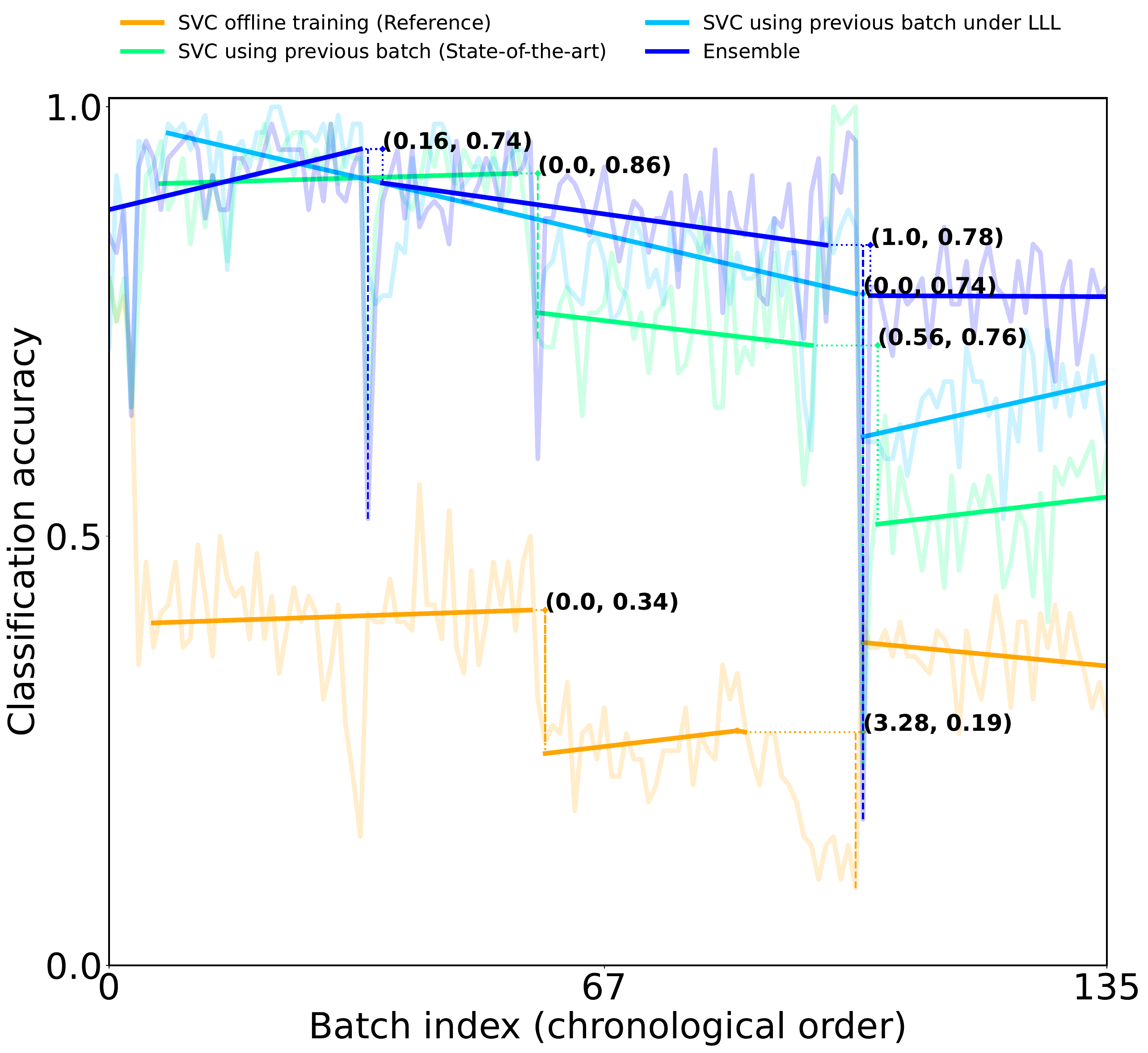}
    \caption{Linear Regression-based dip detection with the (IRM, AUC) metrics annotated as a tuple respectively.}
    \label{fig:example3}
\end{figure}

We will now calculate the IRM and AUC of the performance quality graphs, and \textsc{ResMetric} will use linear regression for dip detection. This method is more robust against noise compared to other dip detection algorithms. This offers a comprehensive assessment of resilience across multiple dimensions. Using linear regression instead is computationally expensive compared to the other dip detection algorithms in \textsc{ResMetric}. In this case study the average computation time of using linear regression was 175 seconds per system. All calculations and results reported in this case study were generated on an HP Elitebook 840 G7 Laptop with i5 Intel 10th generation processor and 16GB memory. The results are visualized in Fig.~\ref{fig:example3}, and there is a difference in dips detected. For example, the LLL system now only has one dip detected.

\subsection{Antifragility analysis using Dip-Dependent Metrics}
Utilizing the resilience metrics R, RR, and RL (adaptive capacity) we will use \proposal to calculate the antifragility of the four systems. We calculated antifragility twice; first using dips detected with the max dips algorithm, and second with linear regression for dip detection. For each system and set up we report \(\alpha_{\bar{u}}\). 

\begin{figure}
    \centering
    \includegraphics[width=0.85\linewidth]{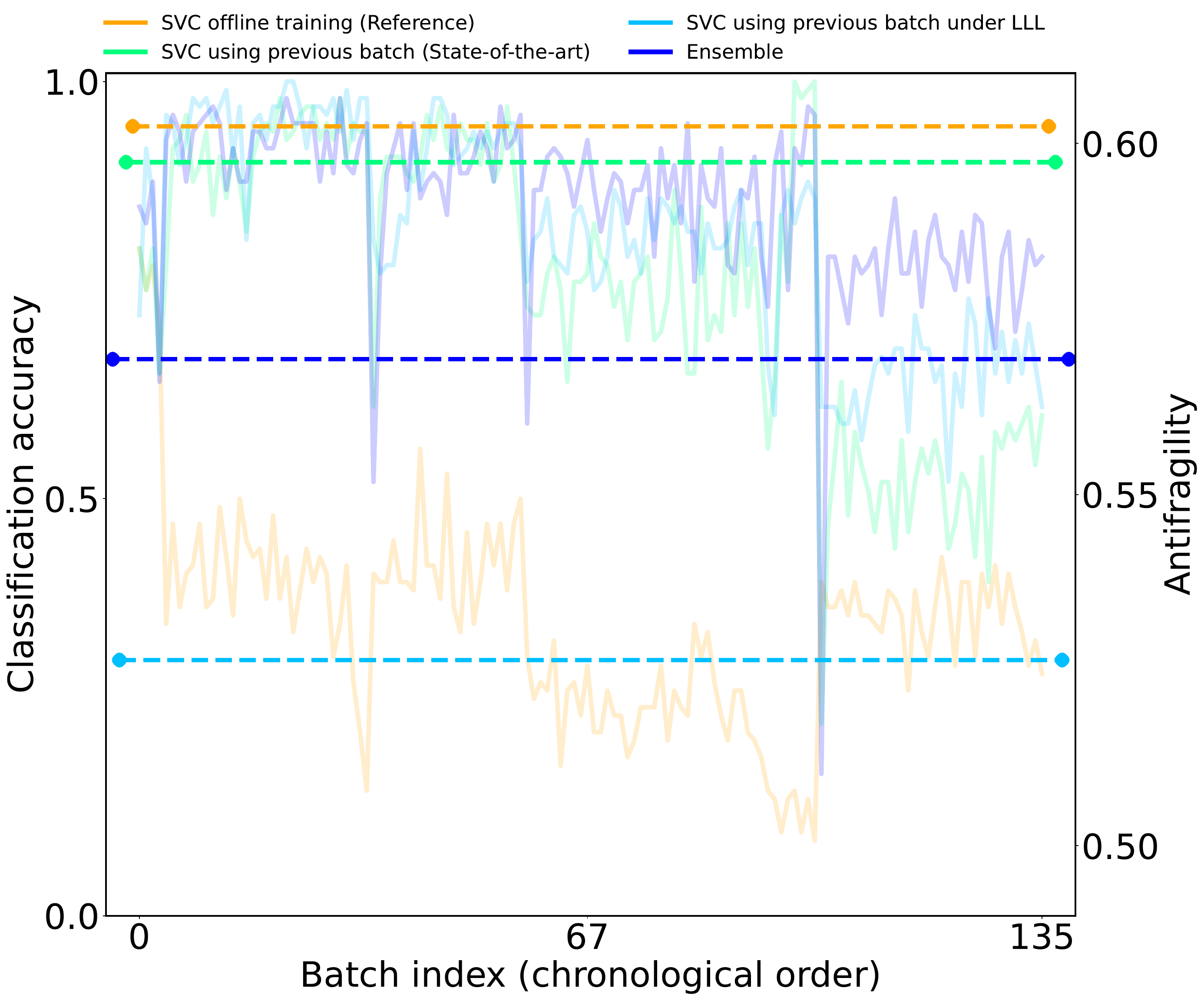}
    \caption{Analysis of \(\alpha_{\bar{u}}\) using the max dips algorithm. Dashed lines are the \(\alpha_{\bar{u}}\) for each system; calculated using R, RR, and RL.}
    \label{fig:example4a}
\end{figure}

\noindent\textbf{Max dips algorithm:} There is a non-expected result of the offline system having the highest \(\alpha_{\bar{u}}\), see Fig.\ref{fig:example4a}. Additionally, the LLL self-adaptive system has the lowest, even the overall performance being comparatively superior to the state-of-the-art and offline, and comparable to the ensemble. 

\begin{figure}[t]
    \centering
    \includegraphics[width=0.85\linewidth]{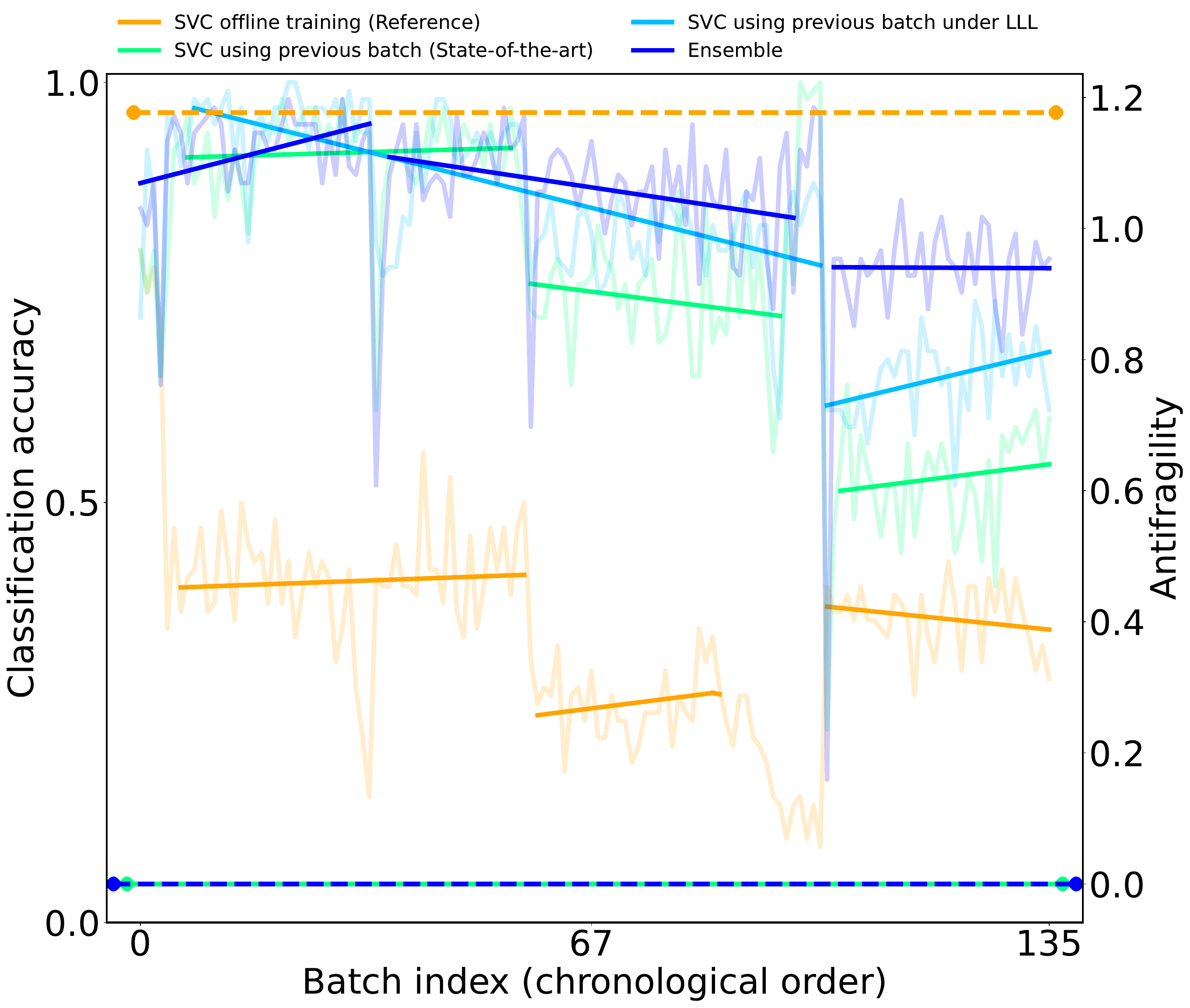}
    \caption{Analysis of \(\alpha_{\bar{u}}\) using the linear regression-based dip detection. Dashed lines are the \(\alpha_{\bar{u}}\) for each system; calculated using R, RR, and RL.}
    \label{fig:example4b}
\end{figure}

\noindent\textbf{Linear regression dip detection:} When switching to the linear regression-based dip detection method fewer dips were detected, see Figs.~\ref{fig:example2},~\ref{fig:example3}. The analysis, see Fig.~\ref{fig:example4b}, differs from the previous antifragility calculations, particularly in cases where only a single dip is identified, which is the case for the LLL system. Consequently, the calculation of \(\alpha_u\) is not feasible for the LLL system. The offline system again reports the highest degree of antifragility and is almost doubled the degree reported when using the max dips algorithm. The other two systems, state-of-the-art and ensemble, have monotonic decreases thus these systems have an overall antifragility degree of 0. 

\balance

\subsection{Discussion}\label{sec:discussion}
From this demonstrative case study, important factors have been highlighted when researchers investigate resilience and antifragility. It is important to be aware of the resilience metrics available and choose appropriately. Further, the choice of dip detection algorithm significantly influences the results. This was particularly evident when analyzing antifragility of the systems, with the LLL system not computable when using linear regression. Additionally, we can observe that while antifragility is a useful concept, it needs to be appropriately weighted when comparing the resilience of systems. Overall, to fully benefit from reporting the degree of antifragility we need to improve our understanding of how best to capture and calculate the definition of antifragility.

\section{Related Work}\label{sec:related}

Yang et al. \cite{yangMicroResVersatileResilience2024} proposed MicroRes, a resilience profiling framework tailored for microservices that injects failures, monitors performance metrics, and evaluates resilience through a degradation dissemination index. MiroRes thus highlights the impact of system failures on user-aware metrics to enhance service reliability. 
With this focus on user-aware metrics, MircoRes specifically involves user interaction, which might be difficult to do in autonomous systems that do not directly engage with users.

Pan et al. \cite{panEmpiricalStudySoftware2023} use five methods (attack surface analysis, static Bayesian network, dynamic Bayesian network, minimal path, and Markov model) to model a system and derive resilience w.r.t. the chosen method.  
However, this approach requires an in-depth understanding of the dynamics and parameters of the system to accurately estimate this system's behavior. This makes it less suitable for scenarios where such detailed prior knowledge is unavailable. In contrast, a model-agnostic approach, as provided by \proposal does not require prior system knowledge, making it more versatile and applicable across diverse system types without deep customization.

Finally, multiple works have calculated and argued over resilience using model checking~\cite{camaraRobustnessDrivenResilienceEvaluation2017, camaraEmpiricalResilienceEvaluation2014, camaraArchitecturebasedResilienceEvaluation2013, camaraEvaluationResilienceSelfadaptive2012}. However, such approaches rely on a specific system model, such as a Discrete-Time Markov Chain, whereas \proposal is model agnostic and can be used for any system (or its model) as long as time series data for the system's performance is available.

\section{Conclusion}\label{sec:conclusion}
This paper introduces \proposal, a comprehensive tool that enables researchers and practitioners alike to visualize different resilience metrics for their systems to enable a meaningful discussion as to which metric is useful for which system and in which context. To this end, \proposal implements resilience metrics found in the literature. 
Additionally, \proposal is designed to allow extensibility, such as additional resilience metrics to be implemented to evaluate the need for new metrics.
As future work, we aim to implement functionality to enable labeling of disruptions to increase effective analysis, and for \proposal to conduct analyses of live systems. We further plan to extend \proposal to utilize additional input formats. 
In particular, we hope that \proposal will enable the community to have meaningful discussions about antifragility. To this end, we supply one proposal for antifragility with \proposal and encourage the community to implement their own proposals. Finally, we demonstrated \textsc{ResMetric} on a system from the literature and showcased that the choice of resilience metric influences its antifragility. Our demonstration visualizes that a naive approach to antifragility might not be sufficient to fully capture the desired properties and more research on antifragility is necessary.

\newpage
\bibliographystyle{IEEEtran}
\bibliography{references}

\end{document}